\documentclass[12pt]{article}

\usepackage{amsmath,amsfonts,bbold,bm}
\usepackage{latexsym}
\usepackage{amssymb}
\usepackage{epsfig,psfrag,cite}
\usepackage{stmaryrd}
\usepackage{multirow,booktabs}

\makeatletter   
\@addtoreset{equation}{section}   
\makeatother

\newcommand{\be}{\begin{equation}}  
\newcommand{\ee}{\end{equation}}  
\newcommand{\bea}{\begin{eqnarray}}  
\newcommand{\eea}{\end{eqnarray}}  
\newcommand{\Tr}{\operatorname{Tr}}
\newcommand{\tr}{\operatorname{tr}}
\newcommand{\str}{\operatorname{str}}
\newcommand{\diag}{\operatorname{diag}}

\renewcommand{\r}[1]{{$\bf #1$}}
\newcommand{\rr}[2]{{$\bf #1_{#2}$}}
\renewcommand{\c}[1]{${\bf \overline{#1}}$}
\newcommand{\cc}[2]{${\bf \overline{#1}_{#2}}$}
\newcommand{\nn}{\nonumber}

\newcommand{\one}{\mathbb{1}}
\newcommand{\s}{\sigma}
\renewcommand{\l}{\lambda}
\renewcommand{\O}{\mathcal O}
\newcommand{\p}{\phantom}
\newcommand{\G}{\mathcal G}
\newcommand{\K}{\mathcal K} 
\newcommand{\loc}{{\rm\, loc}}
\newcommand{\non}{{\rm\, fin}}
\newcommand{\eff}{{\rm eff}}

\begin{document}

\thispagestyle{empty}
\vspace*{.5cm}
\noindent
CERN-PH-TH/2008-113

\vspace*{1.9cm}

\begin{center}
{\Large\bf One-Loop Effective Action in Orbifold Compactifications}
\\[2.3cm]
{\large G.~von Gersdorff}\\[.5cm]
{\it CERN Theory Division, CH 1211 Geneva 23, Switzerland}
\\[.4cm]
{\small\tt (\,gero.gersdorff@cern.ch\,)}
\\[1.3cm]

{\bf Abstract}\end{center} We employ the covariant background
formalism to derive generic expressions for the one-loop effective
action in field theoretic orbifold compactifications. The contribution
of each orbifold sector is given by the effective action of its fixed
torus with a shifted mass matrix. We thus study in detail the
computation of the heat kernel on tori. Our formalism manifestly
separates UV sensitive (local) from UV-insensitive (nonlocal)
renormalization.  To exemplify our methods, we study the effective
potential of 6d gauge theory as well as kinetic terms for
gravitational moduli in 11d supergravity.

\newpage

\section{Introduction}

Orbifolds play a prominent role in both field and string theory
compactifications to four dimensions. They provide the simplest geometries
allowing for four-dimensional (4d) chiral fermions and $\mathcal N=1$
supersymmetry, offer a plethora of symmetry breaking
possibilities, and at the same time possess a high degree of
calculability. They are also particular limits of more general
backgrounds such as Calabi-Yau manifolds, and thus provide a useful
tool to understand these more involved geometries.

Over the last decade or so, field theories with extra dimensions have
become one of the most popular ideas for theories beyond the Standard
Model. Consequently, many papers deal with radiative corrections in
these kind of models
\cite{Hosotani:1983xw,radcorr,vonGersdorff:2002as,Scrucca:2003ra,GrootNibbelink:2003gd,vonGersdorff:2003rq}. An
intriguing feature of these models is that many operators in the 4d
effective action are independent of the UV completion,
as they do not correspond to local counterterms in the higher
dimensional theory, and the UV sensitivity is cut off by the inverse
size of the internal space.  The majority of the literature on
radiative corrections deals with particular orbifolds and applications,
and either sums over the whole Kaluza Klein tower or restrict to the
effective 4d theory.  The latter procedure is very simple but discards
part of the UV completion and thus sacrifices some of the
calculability.  The former procedure grasps the higher dimensional
structure of the theory but can quickly become rather complicated,
especially if one is interested in renormalization of operators beyond
the effective potential.

Without doubt, the most efficient method to calculate the one-loop
effective action (OLEA) is the manifestly covariant method by DeWitt
\cite{DeWitt} and Gilkey \cite{Gilkey:1975iq}. External lines in
Feynman diagrams are traded for a field-dependent mass matrix that is
totally covariant in the background fields. For a noncompact
$d$-dimensional theory, closed expression for the OLEA (up to a fixed
order in the dimension of the operators) can be obtained, that are
valid for particles of any spin. The effort for particular
applications then consists in determining the background dependence of
the mass matrix of the dynamical particles and using this particular
form in the general expressions. Since this can be found by a
linearization of the equations of motion, this method provides an
extremely simple and efficient way to calculate the OLEA.  The central
quantity in the calculation of the effective action is the Schwinger
proper-time propagator, or heat kernel,
\be
K(T)=\exp[-T(-D^2+E)]\,,
\label{hk}
\ee
where $D^2$ is the background covariant d'Alambertian and $E$ the
background dependent mass matrix. The standard evaluation of $K$
proceeds through an expansion in powers of $T$ or, equivalently, in
the dimension of local operators. The goal of this paper is to apply
the covariant background method to orbifold
compactifications.\footnote{Heat kernel techniques have previously
been applied to orbifolds in Ref.~\cite{vonGersdorff:2003dt} in the
context of anomalies. Heat kernel coefficients on boundaries have been
calculated in Refs.~\cite{bd}, see also Ref.~\cite{Vassilevich:2003xt}
and references therein. The case of conical singularities has been
discussed in Refs.~\cite{con}.}  Our formalism avoids the summation
over KK modes altogether and shows the local and nonlocal structure
of these models in a particularly clear way. 

In this paper, we will make the assumption that the fields occuring in
the covariant derivatives as well as the mass matrix $E$ in
Eq.~(\ref{hk}) are independent of the extra dimensinonal coordinates.
Backgrounds of this type allow one to study the effective action of
the light degrees of freedom of most orbifold compactifications, i.e.,
whenever the zero modes have flat profiles.\footnote{Let us stress
though that this assumptions only applies to background fields,
i.e.~we retain the full tower of Kaluza Klein excitations in the
loop.} While this assumption greatly simplifies the results, one loses
some of the invariances inherited from the higher dimensional
theory. As is well known \cite{vonGersdorff:2002as}, gauge invariances
related to normal derivatives lead to a larger invariance group at the
orbifold fixed point than one would naively expect. These additional
symmetries are not manifest in our approach. We will come back to this
issue in the examples and in the conclusions.

The organization of this paper is as follows. In Sec.~\ref{toroidal}
we study the simple case of a toroidal geometry. We will show that the
one-loop trace involves a summation over the torus lattice and propose
a further expansion of the heat kernel coefficients in powers of the
lattice vectors. We explicitly evaluate the coefficients up to
operators of dimension four. In Sec.~\ref{orbifold} we proceed to
orbifold geometries of the type $T^n/Z_N$. We show that each of the
$N$ orbifold sectors generates a contribution that corresponds to the
heat kernel of the sector's fixed torus with a shifted mass matrix. We
also discuss the presence of discrete Wilson lines which modifies the
individual contributions in an interesting way. In Sec.~\ref{examples}
we present two applications of our formalism, the calculation of the
effective potential in 6d $T^2/Z_N$ gauge-Higgs unification, and the
one loop kinetic terms for the gravitational moduli in 11d
supergravity compactified on an orbifold.

\section{Toroidal compactifications}
\label{toroidal}

In this section we would like to analyze the one-loop effective action
of zero modes of the compactification on a torus.\footnote{In this
paper we will be interested only in the effective action of the light
modes, but will take into account all the heavy KK states in the
loop.} As it turns out, the case of the torus provides all the
technical tools for the orbifold compactifications, to be considered
in Sec.~\ref{orbifold}.  The contribution to the OLEA from a generic
field can be written as
\be S_{\rm eff}[A,g,\dots]=-(-)^F\frac{1}{2}\int_0^\infty\frac{dT}{T}\Tr\, \exp[-T(-D^2+E)]\,.
\label{Gamma}
\ee
On the right hand side we have included a field-dependent mass-matrix
$E$ whose form will depend on the particle circulating in the
loop. The derivative is covariant with respect to all gauge and
gravitational symmetries.  We will assume that the zero modes have a
flat profile\footnote{We would like to stress that in principle there
exists no conceptual difficulty in incorporating non-flat profiles in
our formalism. However, the zero modes of many applications do have
flat profiles and we will restrict to these cases in this paper. The
case of nontrivial wave functions, e.g.~warped compactifications,
quasi-localized fields, or massive KK modes is left to future work.}
in the extra dimension, but will take into account an arbitrary
dependence on the 4d coordinates, i.e.~$A_M(x^\mu)$, $g_{MN}(x^\mu)$
etc. This particular background allows one to extract information on
renormalization of operators containing derivatives, such as kinetic
terms. We will frequently use $d$-dimensional covariant quantities
which should be decovariantized at the end.  It is worth noticing that
the inverse propagator can always be cast in the form $-D^2+E$, at
least for a suitable choice of gauge \cite{Vassilevich:2003xt}. The
mass matrices for a fairly generic class of theories are reviewed in
App.~\ref{masses}.
The trace $\Tr$ includes
an integration over spacetime as well as a summation over all internal
indices; in the following we will denote the internal trace by $\tr$.
The exponential in Eq.~(\ref{Gamma}) is called the heat kernel of the
differential operator $-D^2+E$,
\be
K(x,x',T)\equiv \langle x| \exp[-T(-D^2+E)] |x'\rangle\,.
\ee
From its definition it satisfies the differential equation
\be
-\partial_T K(x,x', T)=(-D^2+E)K(x,x',T)\,,
\label{heat}
\ee
and the initial condition
\be
K(x,x',0)=\delta(x-x')\,\one\,.
\label{initial}
\ee

Let us now consider the internal space to be an $n-$dimensional torus
$T^n$ defined by a lattice $\Lambda$ whose elements we will denote by
$\lambda$. The trace can then be written as
\be
\Tr K(T)=\tr \int d^dx\sum_{\lambda\in\Lambda} K(x,x-\lambda,T)\,.
\label{torus1}
\ee
The term in this sum corresponding to $\lambda=0$ will just give rise
to the usual $d$-dimensional one-loop effective action. It describes
particles traveling on closed loops that can be contracted to a point.
On the other hand, the terms with nonzero lattice vectors describe
closed loops that cannot be contracted and, hence, have finite
length. These contributions can never lead to ultraviolet divergent
amplitudes. To see this, it is instructive to directly evaluate
Eq.~(\ref{torus1}) under the assumption that the background fields are
constant. In other words, we calculate the operators in the OLEA that
do not contain any derivatives, i.e~the effective
potential. Introducing a complete set of momentum states\footnote{Note
that the momentum variable is continuous as we are working on the
covering space.} one finds
\begin{align}
K(x,x-\l,T)
&=\int \frac{d^dp}{(2\pi)^{d}}\,
\exp{\left(ip\cdot \lambda -T
[(p-A)^2+E]\right)}\nn\\
&=\frac{1}{(4\pi T)^{d/2}}
\exp{\left(i\lambda\cdot A-\frac{\lambda^2}{4T}-T\,E\right)}\,.
\label{heat2}
\end{align}
Here we have also assumed that $[A_i,A_j]=0$ so that we can
diagonalize the $A_i$ simultaneously and perform the shift in the
momentum variable.  The ultraviolet region of the $T$ integration
corresponds to small $T$ which is thus strongly suppressed for nonzero
$\l$. We will refer to the contributions of non-vanishing $\lambda$ as
nonlocal throughout this paper, while the $\lambda=0$ term is called
local and leads to renormalization of all local $d-$dimensional
operators in the effective action that are compatible with the
symmetries of the theory. All nonlocal operators in the effective
action (including the ones containing derivatives of fields) will come
with an exponential suppression factor as well as the Wilson line
present in Eq.~(\ref{heat2}).\footnote{A perhaps more physical
interpretation can be obtained by representing the propagator or,
equivalently, the heat kernel by a classical path integral. One has to
sum over all closed paths of periodicity $T$ with a weight (action)
given by their geodesic length squared and the associated Wilson line
phase. The non-contractible loops have nonzero length and are always
weighted by $\exp(-\lambda^2/4T)$. This approach has, e.g., been
followed in Ref.~\cite{Brummer:2004xc}.} In the following we will
evaluate the heat kernel on the torus for more general backgrounds.

The standard evaluation of the heat kernel proceeds through an
expansion in powers of $T$ or, equivalently, dimension of the local
operators. To satisfy the initial condition Eq.~(\ref{initial}),
one introduces the following ansatz 
\be 
K(x,x',T)=\frac{1}{(4\pi T)^{d/2}} \bar\Delta(x,x')^{1/2}\,
e^{-\sigma(x,x')/2T} 
\sum_{r\geq 0} T^r a_r(x,x')\,,
\label{ansatzcurved}
\ee
with $a_0(x,x)=\one$.  Here, $\sigma(x,x')$ is the so-called geodesic
biscalar function which, by definition, equals one half the geodesic
distance squared between the points $x$ and $x'$. It satisfies the
following differential equations and initial conditions.
\be
\frac{1}{2}\sigma_{;M}\, \sigma_;^{\, M}=\sigma\,,
\label{biscalar}
\ee
\be
[\sigma_{;M}]=0\,,\qquad[\sigma_{;MN}]=[\sigma_{;M'N'}]=-[\sigma_{;MN'}]=g_{MN}\,,
\ee
where the semicolon denotes covariant differentiation with respect to
unprimed or primed coordinates, and the brackets stand for the
coincidence limit $x'\to x$. The vector $\sigma_;^{\ M}$ has length
equal to that of the geodesic from $x'$ to $x$, is tangent to the
geodesic at $x$ and points in the direction from $x'$ to $x$. The
so-called van Vleck determinant is defined as
\be
\bar \Delta\equiv \det(-\s_{;MN'})\,.
\label{vleck}
\ee
$\bar\Delta$ is a biscalar density with the coincidence limit
$[\bar\Delta]=\det g_{MN}\equiv g$.  The quantities $a_r$ are called
the heat kernel coefficients, they should be considered as bitensors,
gauge transforming at $x$ from the left and at $x'$ from the right. In
particular, the coefficient $a_0$ is the operator of parallel
transport (the Wilson line) connecting the two fibers at $x'$ and $x$
along the geodesic between these two points.
The ansatz Eq.~(\ref{ansatzcurved}) is designed to satisfy the initial
condition Eq.~(\ref{initial}). 
In the standard evaluation of the heat kernel, the quantities needed
are the coincidence limits $[a_r]$, as only these enter in the local
renormalization. There exists various ways to obtain these
quantities, the most straightforward being the recursive procedure
\cite{DeWitt} which we briefly review in App.~\ref{colim}. On
the torus, one needs in addition the periodic coincidence limits
\be 
[a_r]_\l \equiv\lim_{x'\to x-\lambda} a_r(x,x')\,, 
\label{periodic}
\ee
that enter in the nonlocal renormalization. The OLEA can then be
written as
\be
S_{\rm eff}=-(-)^F\int d^dx \sum_{r,\l} \alpha_{d,r}
\frac{[\bar\Delta^\frac{1}{2}]_\l\tr [a_r]_\l}{|\l|^{d-2r}}\,,\qquad 
\alpha_{d,r}=\frac{\Gamma(\tfrac{d}{2}-r)}{2^{2r+1}\pi^\frac{d}{2}}\,.
\label{Stot}
\ee
As expected, only the term with $\l=0$, corresponding to local bulk
renormalization, is UV divergent. Introducing, for
simplicity, a Schwinger cutoff $\exp(-\frac{1}{4\Lambda^2_{UV}T})$ in
Eq.~(\ref{Gamma}), we can write the local and nonlocal
renormalizations in a similar way
\begin{align}
S_{\rm eff,\ loc}&=-(-)^F\int d^dx \sqrt g \sum_r\alpha_{d,r}\label{Sloc}
\,\Lambda_{UV}^{d-2r}\,\tr [a_r]\,,\\
S_{\rm eff, \non}&=-(-)^F\int d^dx \sum_{r,\l\neq0}\alpha_{d,r}
\frac{[\bar\Delta^\frac{1}{2}]_\l\tr [a_r]_\l}{|\l|^{d-2r}}\,.\label{Sfin}
\end{align}
Eqns.~(\ref{Stot}) to (\ref{Sfin}) are only valid for $r<d/2$ because
of infrared (IR) divergences in the $T$ integration that are hidden in
the poles of the Gamma-function present in $\alpha_{d,r}$. Introducing
an IR cutoff $\mu$, one can see that at $r=\frac{d}{2}$ one needs to
make the replacements
\begin{align}
\alpha_{d,\frac{d}{2}}\Lambda_{UV}^0&\to -(4\pi)^{-\frac{d}{2}}
  \log\left(\frac{\mu}{\Lambda_{UV}}\right)\,,\\
\alpha_{d,\frac{d}{2}}|\lambda|^{0}&\to -(4\pi)^{-\frac{d}{2}}\log(\mu |\lambda|)\,.
\end{align}
The IR regulated result valid for arbitrary $r$ is derived in
App.~\ref{zeta}, where we also apply the well-known
zeta-regularization technique to the UV divergences in the local part
of the OLEA.
 
It remains to calculate the periodic coincidence limits, Eq.~(\ref{periodic}).
From the interpretation of $a_0$ as the operator of parallel transport
it is clear that we must have
\be
[a_0]_\l= W(\l)\equiv\exp(i \l^m (A_m+\omega_m))\,.
\label{torus0}
\ee
The Wilson line $W(\l)$ contains both gauge and spin connection
parts, denoted by $A$ and $\omega$ respectively.  For the remaining
coefficients, we expand $a_r(x,x')$ in a covariant Taylor series
around $x'=x$. To this end, we multiply $a_r$ by the Wilson line
$a_0(x',x)$ from the right, so the coefficients are polynomials of
gauge covariant objects at $x$. Then
\be
a_r(x,x')a_0(x,x')^{-1}=[a_r]+[a_{r;M'}]\, \s_;^{M'}
+\frac{1}{2}\,[a_{r;(M'N')}]\, \s_;^{M'}\s_;^{N'}+\cdots
\label{covtaylor}
\ee
Eq.~(\ref{covtaylor}) can easily be proven order by order by
differentiating w.r.t.~$x'$, taking the coincidence limit $x'\to x$,
and using the identities
\be
[\s^{M'}_{\ (N'R'\dots)}]=0\,,\qquad [a_{0;(M'\dots)}]=0\,,
\ee
that can also be proven with the methods reviewed in App.~\ref{colim}.
The result can be expressed as
\be
[a_r]_\l=
[e^{-D_{\l'}}\, a_r] W(\l)
=\biggl([a_r]-[a_{r;\l'}]
+\tfrac{1}{2}[a_{r;\l'\l'}]
+\dots\biggr)W(\l)\,,
\label{torus}
\ee
where $D_{\l'}\equiv\l^mD'_m$. Eq.~(\ref{torus}) is the main
result of this section. 
We have employed a covariant Taylor expansion despite the fact that we
are breaking covariance by the backgrounds.  It proves more efficient
to keep the covariant notation and insert the explicit background at
the end of the calculation. The calculation of the coefficients is
more compact since no distinction is made between the different types
of indices. We will show in App.~\ref{colim} that calculating, e.g.,
$[a_{1;\l'\l'}]$ in the covariant way already provides all necessary
information to find $[a_2]$ etc.  Moreover, we can always use the
classical (tree-level) equations of motion in the on loop corrected
terms, as this is equivalent to a field redefinition
\cite{Burgess:2003jk} (see also Ref.~\cite{Politzer:1980me}). For
instance, we can always replace the $d$-dimensional curvature scalar
$R$ by the trace of the energy momentum tensor, the corresponding
field redefinition being a simple Weyl rescaling.

The mass dimension of the local operators in the parenthesis in
Eq.~(\ref{torus}) is given by $-4+2r+s$ with $s$ the order in the $\l$
expansion. As an example, let us calculate $[a_1]_\l$ and $[a_2]_\l$ up
to dimension four operators, i.e.~we have to evaluate the quantities
$[a_1]$, $[a_2]$, $[a_{1;\l'}]$ and $[a_{1;\l'\l'}]$. The evaluation can be
done in the well-known manner by DeWitts recursive procedure
\cite{DeWitt}. We perform this evaluation in App.~\ref{colim}, one
finds
\begin{align}
[a_1]_\l&=\biggl\{\biggr.
\tfrac{1}{6}R-E-\tfrac{1}{12}R_{;\l}+\tfrac{1}{2}E_{;\l}
    -\tfrac{1}{6}\Omega^M_{\p M \l;M}+\tfrac{1}{40}R_{;\l\l}
    +\tfrac{1}{120} R_{\l\l;\p MM}^{\p{\l\l;}M} \nn\\
&\qquad - \tfrac{1}{90}R^M_{\p M\l} R_{M\l}
    +\tfrac{1}{180}R^{MN}R_{M\l N\l}
    +\tfrac{1}{180}R^{MNL}_{\p{MNL} \l}R_{MNL\l}  -\tfrac{1}{6}E_{;\l\l} \nn\\
&\qquad+\tfrac{1}{24}R^M_{\p M \l}\Omega_{M\l}
    +\tfrac{1}{12}\Omega_{ M\l}\Omega^M_{\p M\l} 
    +\tfrac{1}{12}\Omega^{\p{M\l;}M}_{ M\l;\p M \l}
    +\O(\l^3)\biggl.\biggr\}W(\l)\,,\label{a1}\\
[a_2]_\l&=\biggl\{\biggr.\tfrac{1}{2}(\tfrac{1}{6}R-E)^2
    +\tfrac{1}{6}(\tfrac{1}{5} R-E)^{ M}_{;\ M}-\tfrac{1}{180}R_{MN}R^{MN}\nn\\
&\qquad +\tfrac{1}{180}R^{MNLS}R_{MNLS}+\tfrac{1}{12}\Omega_{MN}\Omega^{MN}
    +\O(\l)\biggl.\biggr\}W(\l)\label{a2}
\,.
\end{align}
Here, $\Omega$ is the field strength of the gauge and spin connections
\be
\Omega_{MN}=[D_M,D_N]=-iF_{MN}+\tfrac{i}{2}\Sigma^{AB}R_{ABMN}\,.
\ee
We will also need the corresponding expansion of the determinant
$\bar\Delta$, 
\be
[\bar\Delta^\frac{1}{2}]_\l=\sqrt g\, (1+\tfrac{1}{12}R_{\l\l}+\O(\l^3))\,.
\ee

The next step is to set $\partial_i=0$ and decovariantize these
expressions. We will leave this step to the explicit examples and end
this section by making a few comments on the form of Eq.~(\ref{a1})
and Eq~(\ref{a2}).
First of all, notice that they contribute to the 4d effective
potential for the 4d scalar zero modes. There are contributions both
from the Wilson line as well as the prefactors $[A_i,A_j]^2$,
$[A_i,E]$ etc. However, restricting to the tree level flat directions,
$[A_i,A_j]=0$ (i.e.~the moduli space), we see that many contributions
vanish. This is the expansion resulting from Eq.~(\ref{heat2}). If, in
addition, we assume that $A_i$ and $E$ commute, then the only constant
terms left in $a_r$ result from the expansion of the field dependent
mass suppression, $e^{-T E}$.

The next comment concerns the IR divergences of the double
expansion in $T$ and $\l$. The integration over $T$ is IR divergent
when $2r\geq d$, as is evident from the presence of the $\Gamma$ function
in Eq.~(\ref{Stot}). On the other hand, the summation over $\l$
produces IR divergences once the dimension of the operator exceeds
four. It is worth noticing that if the matrix $E$ is positive definite
its smallest eigenvalue provides an effective $d$-dimensional IR
cutoff, in which case one gets a good approximation if one includes in
the summation over $\Lambda$ only the terms with small $|\l|$. This
corresponds to closed loops that only wind a few times around the
torus, which dominate the IR behavior.


%
%

\section{Orbifold Compactification}
\label{orbifold}

In order to obtain phenomenologically more interesting models, we
would like to orbifold the toroidal geometries considered in the
previous section. The $Z_N$ orbifold\footnote{Here we consider only
orbifolds with one $Z_N$ factor. The generalization to several
factors, or even nonabelian groups, is straightforward. We also
rewrite orbifolds that involve non-integral lattice shifts as integral
shifts with discrete Wilson lines.} is constructed by identifying
points that are related by a rotation of the torus:
\be
x\sim P^k\,x-\l\,,\qquad \l\in\Lambda\,,\qquad P^N=1\,.
\label{spacegroup}
\ee
This operation is well defined on the torus only if the $Z_N$ action
defines an automorphism of the torus lattice, i.e., maps $\Lambda$ to
itself. This property, also known as the crystallographic principle,
greatly restricts the allowed lattices and values for $N$. These are
well known and classified for the dimensions most interesting for
phenomenological applications, see e.g.~Ref.~\cite{Bailin:1999nk} for
the 10d case.  The $Z_N$ group acting as rotations is known as the
point group $G$, while the one generated by both lattice translations
and rotations is called the space group $S$. We can decompose each
$g\in S$ as in Eq.~(\ref{spacegroup}) and accordingly write
$g=(k,\l)$. The space group is represented on the fields as
\be
\phi(g x)= W_0(\l)\,(P_L\otimes P_\mathcal G)^k \phi(x)\,,
\label{spacegroupfields}
\ee
where $P_L$ is the representation of $P$ on the Lorentz group and
$P_\mathcal G$ acts on all internal indices (in particular the gauge
group $\mathcal G$). We have also included a discrete Wilson line
$W_0$. Discrete Wilson lines commute with $P_\mathcal G$ and satisfy
\be
W_0(P\l)=W_0(\l)\,,\qquad W_0^N=\one\,.
\label{dwl}
\ee
For reasons of clarity we will present the calculations of this
section for $W_0=\one$ and only give the relevant results for
nontrivial $W_0$ at the end.

\begin{figure}[thb]
\psfrag{x}{$x_\parallel$}
\psfrag{y}{$x_\perp$}
\psfrag{R}{$\mathbb R^4$}
\psfrag{L}{$\mathbb R^{d_\parallel-4},\ \Lambda_\parallel$}
\psfrag{K}{$\mathbb R^{d_\perp},\ \Lambda_\perp$}
\includegraphics[width=\linewidth]{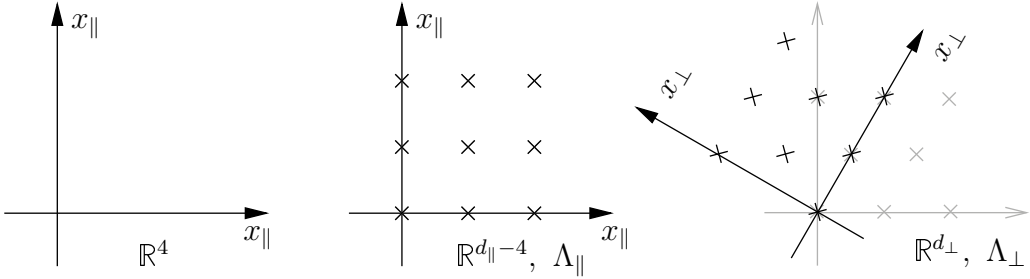}
\caption{In each $k$ sector of the orbifold, the coordinates split
into fixed ($x_\parallel$) and rotated ($x_\perp$) under the action of
$P^k$. The crosses indicate the lattice $\Lambda$ of the underlying
torus, which also splits into the direct sum
$\Lambda=\Lambda_\parallel+\Lambda_\perp$. Notice that either torus
can be trivial for particular sectors.}
\label{not}
\end{figure}

In order to calculate the effective action in the orbifolded theory,
we make use of the fact that any field satisfying the point group
constraint 
\be
\phi_{orb}(Px)=(P_L\otimes P_\G)\phi(x)_{\rm orb}
\ee
can be obtained from the fields on the torus by applying the linear
projection
\be
\phi_{\rm orb}(x)=\frac{1}{N}\sum_{k=0}^{N-1}(P_L\otimes P_\G)^{-k} 
\phi_{\rm tor}(P^{k}x)
\ee
on any torus field. Consequently, we can evaluate the trace on the
orbifold as\footnote{The projection method was first developed for the
codimension-one case \cite{bd}.}
\be
\Tr K(T)=\frac{1}{N}\int dx\, \tr_{\rm }K(x,P^kx-\l)(P_L^\dagger
\otimes P^\dagger _G)^{-k}\,.
\ee
Following the notation of Ref.~\cite{GrootNibbelink:2003gd}, for a
given point group element $P^k$ we split the covering space according
to $\mathbb R^d=\mathbb R^{d_\parallel}\oplus \mathbb R^{d_\perp}$,
where by definition the $d_\parallel$ coordinates $x_\parallel$ are
left fixed by $P^k$, see the illustration in Fig.~\ref{not}. This
splitting obviously depends on $k$, in order to avoid a cumbersome
notation such as $x_{k,\parallel}$ etc.~we will omit the index $k$
when no confusion can arise. In the same way we split the torus
$\Lambda=\Lambda_\parallel+\Lambda_\perp$.  For the orbifold we need
to evaluate the matrix element
\be
K(x,P^kx-\l)= \langle x|\exp\left(-T[-D^2+E]\right)|P^k x-\l\rangle\,.
\ee
Using the splitting just introduced, one finds
\begin{multline}
 K(x,P^kx-\l)
=
\int 
dp_\perp
\exp \biggl(ip_\perp(P^k-1)\left[x_\perp-x_f(\lambda_\perp)\right]\biggr)
\times\\\times
\langle x_\parallel|\exp\biggl( -T\left[-D_\parallel^2+(p_\perp-A_\perp-\omega_\perp)^2+E\right]\biggr)
|x_\parallel-\lambda_\parallel\rangle \,,
\end{multline}
where we have used that the nonsingular matrix $P^k-1$ provides a
one-to-one map from the set of fixed points on the transverse space to
the lattice vectors in $\Lambda_\perp$.
%
The next thing we would like to do is to perform the trace over the
transverse torus, i.e.~we would like to perform the integration/summation
\be
\Tr_\perp=\sum_{\Lambda_\perp}\int_{\mathcal F_\perp}  dx_\perp\  \,,
\ee
where the integration is over the fundamental domain $\mathcal
F_\perp$ of the torus. We now replace the sum over $\Lambda_\perp$ by
the sum over all fixed points in the covering space, again by virtue
of the map.  We then can write
\begin{multline}
\Tr_\perp \exp\biggl( ip_\perp(P^k-1)\left[x_\perp-x_f(\l_\perp)\right]\biggr)\\
=\sum_{x_f}\int_{\mathcal F_\perp}  dx_\perp
\exp(\dots)
=\sum_{x_f\in \mathcal F_\perp}\int  dx_\perp
\exp(\dots)
\end{multline}
where the integration is now over the whole covering space whereas the
summation over the fixed points is restricted to the fundamental
domain. From now on all summations over fixed points are implicitly
assumed to be only over $\mathcal F_\perp$.  The integration over
$x_\perp$ gives $|\det(1-P^k)|^{-1}\delta(p)$. According to Lefshetz'
formula, the determinant equals the number of fixed points in
$\mathcal F_\perp$, leading to
\be
\Tr_\perp  K(T)
= \exp{\left(-T
[-D_\parallel^2+E+(A_\perp+\omega_\perp)^2]\right)}\equiv  K_{\parallel}(T)\,.
\label{heat3}
\ee
The final result on the orbifold without discrete Wilson lines is thus
\be
S_{\rm eff}=-(-)^F\frac{1}{2N}\int \frac{dT}{T}\sum^{N-1}_{k=0}  
\Tr (P_\G\otimes P_L)^k  K_{\parallel}(T)\,.
\label{orbnodw}
\ee
The trace in Eq.~(\ref{orbnodw}) includes an integration
over the $d_\parallel$ dimensions $x_\parallel$ as well as a summation
over the lattice $\Lambda_\parallel$ of the fixed torus of the $k^{th}$
sector.  One concludes that the renormalization of the $k^{th}$ orbifold
sector (i.e., the $k^{th}$ term in the sum) is localized on the
corresponding fixed torus. Moreover, the UV sensitive contribution,
$\l_\parallel=0$, is the local renormalization at the fixed points.
The evaluation of Eq.~(\ref{orbnodw}) now proceeds precisely as described in
Sec.~\ref{toroidal} in $d=d_\parallel$ dimensions, the only difference
being the shifted mass matrix and the orbifold twists inside the
trace. In particular, Eqns.~(\ref{torus0}), (\ref{a1}), and (\ref{a2})
are still valid.  Notice, however, that the original mass matrix $E$
is still the one obtained in the $d$-dimensional theory. For instance,
a non-minimally coupled scalar has $E=\eta R_d$, the $d$ dimensional
curvature scalar, the mass matrix for a vector particle is still a
$d\times d$ matrix etc.

In case there are discrete Wilson lines, we first perform the splitting
\be
W_0(\l)=W_0(\l_\perp+\l_\parallel)=W_0(\l_\perp)W_0(\l_\parallel)\,.
\ee
The Wilson line $W_0(\l_\parallel)$ just multiplies the background
(continuous) Wilson line $W(\l)$ occurring in the periodic coincidence
limit of the heat kernel coefficients, i.e.~Eqns.~(\ref{torus0}),
(\ref{a1}) and (\ref{a2}).  To take into account the effect of
the orthogonal Wilson line, one has to introduce the following matrix
in the trace in Eq.~(\ref{orbnodw})
\be
Q_\perp= |\det(1-P^k)|^{-1} \sum_{x_{f,k}} W_0(\lambda_\perp(x_{f,k}))\,,
\label{proj}
\ee
leading to~\footnote{In deriving Eq.~(\ref{orbdw}) one has to use Eq.~(\ref{dwl}).}
\be
S_{\rm eff}=-(-)^F\frac{1}{2N}\int \frac{dT}{T}\sum^{N-1}_{k=0} 
\Tr (P_\G\otimes P_L)^k
Q_{\perp}  K_{\parallel}(T)\,.
\label{orbdw}
\ee
As emphasized earlier, the splitting into $x_\parallel$ and $x_\perp$
depends on the orbifold sector (i.e.~on $k$), and, as a consequence,
the same holds true for the quantities $Q_\perp$ and $K_\parallel$.
One can interpret this result by noting that the quantity $Q_\perp$ is
nothing but the projector onto zero modes on the transverse torus
defined by the lattice $\Lambda_{\perp}$, i.e., the zero modes that
would be obtained from compactification on the torus $\Lambda_{\perp}$
with the discrete Wilson lines $W_0(\l_\perp)$ present.  These
projectors actually take very simple forms in concrete examples, as
the possible Wilson lines are very restricted. We will give the
explicit forms of $Q_\perp$ for the $T^2/Z_N$ orbifolds in
Sec.~\ref{examples}.

Let us emphasize an important point. The contribution with
$\l_\parallel=0$ corresponds to a local renormalization at the fixed
points of the $k$-sector of the orbifold. As expected, these are UV
divergent and should respect all symmetries preserved at the fixed
point. As discussed in the literature \cite{vonGersdorff:2002as}, the
gauge symmetries actually further constrain the allowed operators
because of shift symmetries related to normal derivatives. These
remnant gauge symmetries are not manifest in our formalism due to the
fact that we have only considered backgrounds with vanishing normal
derivatives.\footnote{We will come back to this issue in
Sec.\ref{examples}.} With some effort one can set up a fully covariant
heat kernel expansion that manifestly displays the surviving
symmetries at a given fixed point. However, the resulting formulae are
considerably more involved and we will leave this to future research. 

Finally, localized matter (twisted sectors) can appear on the fixed
tori. In the trivial case of a 4d fixed point, their contribution is
just the usual 4d one
\be
S_{\eff}^{\rm twisted}=-(-)^F\frac{1}{2}
\int d^4x\,\tr\,K_{4d}(x,x,T)\,.
\ee
For higher dimensional fixed points, the geometry seen by these fields
is again an orbifold, of dimension $d'<d$ and order $N'<N$ which can
be treated as before.

\section{Examples}
\label{examples}
\subsection{6d Gauge Theory}

As our first example for the use of our methods, we consider the
effective action in 6d gauge-Higgs unification models on the
orbifold $T^2/Z_N$.  In these type of models, the bulk gauge group
$\mathcal G$ is broken to a subgroup $\mathcal H$ by the orbifold
twist $P_\mathcal G$. The Higgs for the further breaking of $\mathcal
H$ then resides in the $A_{4,5}$ components of the gauge field
belonging to the coset $\mathcal G/\mathcal H$.  The tree-level
potential derives from the $F_{MN}^2$ kinetic term in the action. It
is important to distinguish two kinds of 4d scalar fields resulting
from the compactification: generic massless ones (orbifold invariant
states) and flat directions (a subset of the zero modes corresponding
to the condition $[A_i,A_j]=0$). Not all zero modes correspond to flat
directions.

We will consider pure gauge theory and calculate the contribution from
gauge and ghost loops, the corresponding mass matrices are given in
App.~\ref{masses}. In flat gravitation
background they read
\be
E_{1,MN}=2i F_{MN}\,,\qquad E_{1,gh}=0\,.
\ee
The result for the $k=0$ sector can be read off from
Eqns.~(\ref{Sloc}) to (\ref{torus0}) as well as (\ref{a1}) and (\ref{a2}). 
\begin{align}
S^{k=0}_{\eff,\loc}&=
-\frac{1}{N}\int d^6x \biggl\{\biggr.
\frac{4\dim(\G)\, \Lambda_{UV}^6}{\pi^3}
+\frac{5C_2(\G)\,\Lambda_{UV}^2}{96\pi^3}F_{MN}^aF^{a,MN}
\biggl.\biggr\}\label{6d0}\\
S^{k=0}_{\rm eff, \non}&=-\frac{1}{N}\int d^6x\sum_{\l\neq0} \tr \biggl\{\biggr.
\frac{4}{\pi^3|\l^6|}W(\l)
 +\frac{1}{12\pi^3|\l^4|}W(\l)\bigl[\bigr.iF^M_{\p M\l;M}\nn\\
&\qquad -\tfrac{1}{2}F_{M\l}F^{M\l}
-\tfrac{i}{2}F_{M\l;\p M\l}^{\p{M\l;}M}\bigr.\bigr]
+\frac{5}{96\pi^3|\l|^{-2}} W(\l)F_{MN}F^{MN}
\biggl.\biggr\}\,,\label{6d1}
\end{align}
where the integration is over the volume of the torus, the trace in
the adjoint representation, and we recall from Sec.~\ref{toroidal} our
shorthand notation $X_\l\equiv\l^iX_i$.  Eq.~(\ref{6d0}) is the
renormalization of the bulk cosmological constant and the bulk gauge
kinetic term.  Eq.~(\ref{6d1}) contains the Hosotani potential
\cite{Hosotani:1983xw}, kinetic terms for $A_{\mu}$, as well as
potential and kinetic terms for $A_{4,5}$.

The contributions from the sectors with $k\neq 0$ correspond to the
renormalizations at the fixed points. The fixed points are four
dimensional and contain no further toroidal dimensions, so there is
only a local renormalization.  We define the orbifold action on the
coordinates to be a counterclockwise rotation of angle $2\pi k/N$. The
action on the gauge fields thus reads
\be
P_L^k=\left(\begin{array}{ccc}
\one_4 &&\\
&c_k&s_k\\
&-s_k&c_k
\end{array}
\right)\,,\qquad c_k=\cos\left(\tfrac{2\pi k}{N}\right)\,,
   \qquad s_k=\sin\left(\tfrac{2\pi k}{N}\right)\,.
\ee
The gauge loop gives
\begin{multline}
  S_{\eff,\loc}^{k\neq0,{\rm vector}}=
-\frac{1}{N}\int d^4x\ \tr\, \biggl\{\biggr.(P_L\otimes P_\mathcal G)^k\times\\
\times\biggl(\biggr.
\frac{\Lambda_{UV}^4}{2\pi^2} 
-\frac{\Lambda_{UV}^2}{8\pi^2} E_1'
-\frac{\log \frac{\mu}{\Lambda_{UV}}}{192\pi^2}\left[ 6 E_1'^2- F_{\mu\nu}F^{\mu\nu}\right]
\biggl.\biggr)
\biggl.\biggr\}\,,
\label{vcon}
\end{multline}
where the shifted mass
matrix $E_1'\equiv E_1+A_\perp^2$ reads
\be
E'_{1,MN}=2iF_{MN}+A_kA^{k}\delta_{MN}\,,
\ee
with $A_k$ and $F_{MN}$ considered as matrices in the adjoint
representation.  The ghosts correspond to two scalars. Their
contribution is thus obtained by setting $P_L=1$ in Eq.~(\ref{vcon}),
multiplying by $-2$, and using the mass matrix
\be
E_{1,gh}'=A_kA^k\,.
\ee
Adding up the contribution of the gauge fields and the ghosts and
performing the trace over the Lorentz indices one obtains
\begin{multline}
  S_{\eff,\loc}^{k\neq0}=
-\frac{1}{N}\int d^4x\ \tr\, \biggl\{\biggr.P_\G^k
\biggl(\biggr.
\frac{(c_k+1)\Lambda_{UV}^4}{\pi^2}
+\frac{(c_k-3)(1+c_k-is_k)\Lambda_{UV}^2 }{8\pi^2} BB^\dagger\\
-\frac{(c_k+5)(1+c_k-is_k)\log \frac{\mu}{\Lambda_{UV}}}{64\pi^2} 
   BB^\dagger B B^\dagger
+\frac{(3c_k-1)\log \frac{\mu}{\Lambda_{UV}}}{32\pi^2} B^\dagger B^{2}B^\dagger
\\
-\frac{\log \frac{\mu}{\Lambda_{UV}}}{4\pi^2}
 F_{\mu i}F^{\mu i}
+\frac{(c_k-11)\log \frac{\mu}{\Lambda_{UV}}}{96\pi^2}
 F_{\mu\nu}F^{\mu\nu}
\biggl.\biggr)
\biggl.\biggr\}\,,
\label{locres}
\end{multline}
where we have defined $B=A_4+iA_5$ and made use of the fact that the
orbifold boundary condition implies $BP^k_\G=P^k_\G B (c_k-is_k)$.  As
stressed earlier, the result is not covariant w.r.t.~the remnant gauge
symmetry related to the normal derivatives
\cite{vonGersdorff:2002as}. This is obviously so, as we have
explicitely set to zero all normal derivatives in order to obtain the
simple result in Eq.~(\ref{orbnodw}). In the present case it is,
however, easy to reconstruct the covariant structure as follows. The
potential should result from the following operators
\cite{vonGersdorff:2002as}
\be
\tr P_\G^k F_{45}\,,\qquad \tr P_\G^k (F_{45})^2\,,\qquad \tr P^k_\G F_{45;i}^{\p{45;i}i}.
\label{inv}
\ee
Using again the orbifold boundary conditions, we can write
\begin{eqnarray}
\tr P_\G^k F_{45}&=&\frac{1-c_k+is_k}{2}\tr P_\G^k BB^\dagger\\
\tr P_\G^k(F_{45})^2&=&\frac{1+c_k-is_k}{4}\tr P_\G^k BB^\dagger B B^\dagger
    -\frac{1}{2}\tr P_\G^kB^\dagger B^2B^\dagger\\
\tr P^k_\G F_{45;i}^{\p{45;i}i}&=&-(1-c_k+is_k)\tr P_\G^k BB^\dagger B B^\dagger
+i s_k\tr P_\G^kB^\dagger B^2B^\dagger\ \   
\end{eqnarray}
These relations can clearly be inverted and used to replace the
operators occuring in Eq.~(\ref{locres}) by the covariant ones.
One finds:
\begin{multline}
  S_{\eff,\loc}^{k\neq0}=
-\frac{1}{N}\int d^4x\ \tr\, \biggl\{\biggr.P_\mathcal G^k
\biggl(\biggr.
\frac{\Lambda_{UV}^4(c_k+1)}{\pi^2}
-i\,\frac{\Lambda_{UV}^2 (c_k-3)(c_k+1)}{4\pi^2 s_k} F_{45}\\
-i\,\frac{\log \frac{\mu}{\Lambda_{UV}}(c_k-3)(c_k+1)}{16\pi^2(c_k-1)s_k} 
   F_{45;i}^{\p{45;i}i} 
-\frac{\log \frac{\mu}{\Lambda_{UV}}(c_k^2+7)}{16\pi^2(c_k-1)}(F_{45})^2
\\
-\frac{\log \frac{\mu}{\Lambda_{UV}}}{4\pi^2}
 F_{\mu i}F^{\mu i}
+\frac{\log \frac{\mu}{\Lambda_{UV}}(c_k-11)}{96\pi^2}
 F_{\mu\nu}F^{\mu\nu}
\biggl.\biggr)
\biggl.\biggr\}\,.
\end{multline}
%
It is, however, not clear if this procedure can be generalized to
higher dimensional fixed points and gravitational symmetries. First,
one would need to find an independent set of covariant operators, as
in Eq.~(\ref{inv}), suitable for the surviving symmetries at the fixed
point. Given this set, it is not clear whether there is a one-to-one
correspondance to the operators obtained with the simpler background
constant in the normal directions. We believe that the better approach
is to directly compute a fully covariant heat kernel expansion that
manifestly displays all gauge symmetries inherited from the higher
dimensional theory. This approach will be presented elsewhere
\cite{future}, along the lines presented in Sec.~\ref{conclusions}.

Including discrete Wilson lines is simple.  First note that
Eq.~(\ref{dwl}) implies that the two discrete Wilson lines have to be
of order 2, 3, 2 and 1 for $N=2,\ 3,\ 4$ and 6 respectively, they also
have to coincide for $N\neq 2$.  Eq.~(\ref{6d0}) remains unaltered in
the presence of discrete Wilson lines, while in Eq.~(\ref{6d1}) the
background Wilson lines become multiplied by $W_0(\l)$. Finally, the
localized renormalizations now include the projectors $Q_\perp$. The
explicit forms of these projectors are
\be
Q_\perp^{N=2}=\frac{1}{4}(\one+W_{0,1})(1+W_{0,2})\,,\quad
Q_\perp^{N=3}=\frac{1}{3}(\one+W_0+W_0^2)\,,
\ee
\be
Q_\perp^{N=4}=\frac{1}{2}(\one+W_0)\,.
\ee
For $N=6$ one necessarily has $W_0=\one$ and hence the projector is
trivial.


\subsection{11d Supergravity}

In this section we would like to calculate the one-loop corrections to
the kinetic terms of the gravitational moduli in an orbifold
compactification of 11d supergravity. This is an important quantity as
it largely determines the one-loop renormalization of the K\"ahler
potential which, in turn, determines the scalar potential once
supersymmetry is broken. The perturbative scalar potential is of high
relevance due to the large number of moduli that need to be stabilized
in these models.

We will consider an $\mathcal N=1$ compactification on the space
$T^6/Z^3\times S^1/Z_2$. The $Z_3$ action is given by the $U(3)\subset
SO(6)$ preserving shift vector $\phi=(1,1,-2)$.  The $T^6$ complex
torus coordinates transform under $Z_3$ as
\be
z_1\to e^{2\pi i/3}z_1\,,\qquad z_2\to e^{2\pi i/3}z_2\,,\qquad 
z_3\to e^{-4\pi i/3}z_3\,.
\ee
The $Z_2$ action is given by $x^{10}\to-x^{10}$. To keep a compact
notation we will write everything in terms of $Z_6$ generated by
$P=P_{Z_2}(P_{Z_3})^{-1}$.

The field content of our 11d theory is a bulk supergravity multiplet
consisting of the graviton, a Majorana gravitino and an antisymmetric
three-form $B$. To cancel localized anomalies, we would introduce
$E_8$ and $E_8'$ gauge multiplets at the two fixed points of the
$S^1/Z_2$ orbifold \cite{Horava:1995qa}. In this paper, we will restrict ourselves to the
supergravity sector only.  The parities of these fields are as
follows. Each vectorial index on the metric, the gravitino, and the
three-form transform as the coordinates. There is
an additional overall minus sign for the $B$-field w.r.t.~$Z_2$,
i.e.~$B_{\mu\nu\rho}$ has negative parity under reflection of
$x^{10}$. Finally, the spinor indices transform with $\gamma^{10}$
under $Z_2$.\footnote{Recall that we use Euclidean conventions with
$\{\gamma^A,\gamma^B\}=2\delta^{AB}$.}  This assignment results in the
orbifold twists displayed in Tabs.~\ref{parities} to \ref{paritiesf}.

\begin{table}[t]
\begin{center}
\begin{tabular}{|cc||c|cc|cc|}
%
\hline
&&&&&&\\[-12pt]
&
&$h=2$&\multicolumn{2}{c|}{$h=1$}&\multicolumn{2}{c|}{$h=0$}
\\
\hline
&&&&&&\\[-12pt]
\multirow{2}{.45cm}{$Z_2$}
&$+1$      &\r1 &\r6&     &\r{20'}+$2\times$\r1   &       \\
&$-1$      &    &   &\r1  &                       &\r6    \\
\hline
&&&&&&\\[-12pt]
\multirow{3}{.45cm}{$Z_3$}
&$\theta^0$&\r{1_0} &          &\r{1_0}&\r{8_0}+2$\times$\r{1_0}&           \\
&$\theta^1$&        &\rr{3}{1}&       &\cc{6}{-2}              &\rr{3}{1} \\
&$\theta^2$&        &\cc{3}{-1} &       &\rr{6}{2}             &\cc{3}{-1}  \\
\hline
\end{tabular}
\caption{The orbifold parities of the metric (\rr{44}{}) of the
supergravity multiplet. Here, $h$ denotes the 4d helicity and
$\theta=e^{2\pi i/3}$.  For the $Z_2$ parities, we label fields by
$SO(6)$ irreducible representations. For $Z_3$ we write fields in
terms of representations of the surviving $U(3)$ with the $U(1)$
generator normalized as $Q=\Sigma^{45}+\Sigma^{67}+\Sigma^{89}$. }
\label{parities}
\end{center}
\end{table}

\begin{table}[t]
\begin{center}
\begin{tabular}{|cc||cc|cc|}
\hline
&&&&&\\[-12pt]
&
&\multicolumn{2}{c|}{$h=1$}&\multicolumn{2}{c|}{$h=0$}
\\
\hline
&&&&&\\[-12pt]
\multirow{2}{.45cm}{$Z_2$}
&$+1$      &\r6   &       &\r{15}+\r1  & \\
&$-1$      &        &\r{15} &            &\r{20'}+\r6 \\
\hline
&&&&&\\[-12pt]
\multirow{3}{.45cm}{$Z_3$}
&$\theta^0$&          &\r{1_0}+\r{8_0}&\r{8_0}+2$\times$\r{1_0} &\r{8_0}  \\
&$\theta^1$&\rr{3}{1} &\rr{3}{-2}      &\rr{3}{-2}     &\cc{6}{-2} \\
&$\theta^2$&\cc{3}{-1}  &\cc{3}{2}     & \cc{3}{2}   &\rr{6}{2} \\
\hline
\end{tabular}
\caption{The orbifold parities of the 3-form (\rr{84}{}) of the
supergravity multiplet. See explanations below Tab.~\ref{parities}. }
\label{paritiesp}
\end{center}
\end{table}

\begin{table}[t]
\begin{center}
\begin{tabular}{|cc||cc|cccc|}
\hline
&&&&&&&\\[-12pt]
%
&&\multicolumn{2}{c|}{$h=3/2$}
&\multicolumn{4}{c|}{$h=1/2$}\\
\hline
&&&&&&&\\[-12pt]
\multirow{2}{.45cm}{$Z_2$}
&$+1$&\r4&      &2$\times$\c4&            &\c{20}&\\
&$-1$&   &\c4   &            &$2\times$\r4&      &\r{20}\\
\hline
&&&&&&&\\[-12pt]
\multirow{3}{.45cm}{$Z_3$}
&$\theta^0$&\rr{1}{3/2}&\rr{1}{-3/2}  &2$\times$\rr{1}{-3/2}&$2\times$\rr{1}{3/2}&\rr{8}{-3/2}    &\rr{8}{3/2}\\
&$\theta^1$&\rr{3}{-1/2} &           &                  &$2\times$\rr{3}{-1/2} &\rr{3}{5/2}     &\rr{3}{-1/2}+\cc{6}{-1/2}\\
&$\theta^2$&          &\cc{3}{1/2} &2$\times$\cc{3}{1/2}&                  &\cc{3}{1/2}+\rr{6}{1/2} &\cc{3}{-5/2}\\
\hline
\end{tabular}
\caption{The orbifold parities of the gravitino (\rr{128}{}) in the supergravity
multiplet. See explanations below Tab.~\ref{parities}.}
\label{paritiesf}
\end{center}
\end{table}

We will focus on the following background:
\be 
g_{MN}=\diag\left(g_{\mu\nu},\ \rho_1^2,\ \rho_1^2,\ \rho_2^2,\ \rho_2^2,\ \rho_3^2,\ \rho_3^2,\ \sigma^2\right)\,,\qquad 
B_{MNR}=0\,,\label{bg}
\ee
where all fields $g_{\mu\nu}$, $\rho_I$ and $\sigma$  are
assumed to be independent of the internal coordinates $x^i$.  This
does not cover all zero modes in the supergravity multiplet: From
Tab.~\ref{parities}, \ref{paritiesp} and \ref{paritiesf} for instance
one can see that the $\mathcal N=1$ chiral superfields come in
$SU(3)$-multiplets: There are two singlets as well as one octet. The
two singlets correspond to the two volume moduli of $T^6$ and $S^1$
respectively, whereas the octet describes the precise shape of the
$T^6$ torus. However, it turns out that it is sufficient to consider
the simplified background, Eqns.~(\ref{bg}), and reconstruct the full
kinetic terms by $SU(3)$ invariance.
Before doing any detailed calculation, let us summarize the different
places where contributions to the kinetic terms of the gravitational
moduli can arise. After restricting to the background Eqns.~(\ref{bg})
the heat kernel coefficients quadratic in the 4d derivatives
are
\begin{align}
[\bar\Delta^\frac{1}{2}]_\l&=\frac{1}{12}\sqrt {g}\, R_{\l\l}\,,\label{con0}\\
[a_0]_\l&=-\frac{1}{2} (\lambda \cdot\omega_\parallel)^2\,,\label{con1}\\
[a_1]_\l&=\frac{1}{6}R-E-\omega_{\perp}^2\label{con2}\,.
\end{align}
In Eq.~(\ref{con1}) we have expanded the Wilson line to second order
in the spin connection, which is linear in the 4d derivative.  We now
parametrize the full result as follows. Let us combine the sectors
according to their fixed tori. There are thus 4 sectors, corresponding
to the $Z_N$ elements with $N=1,2,3,6$. They possess fixed tori $T^7$,
$T^6$, $T^1$, $T^0$ and have $\mathcal N=8,4,2,1$ supersymmetry
respectively. Then the one-loop kinetic terms are
\be
\K=\mathop{\sum_{d=11,10,5,4}}_{r=0,1}(\K_d^{r,\loc}+\K_d^{r,\non})
\label{cont}
\ee
where the finite nonlocal contributions result from the terms with
$\l\in\Lambda_\parallel$ nonvanishing, and the local UV-sensitive ones
from the term with $\l=0$.  Recall that $\Lambda_\parallel$ by
definition is the lattice of the fixed torus associated to each
orbifold sector.  Clearly, $\K_d^{0,\loc}=0$ as this contribution
occurs only for nonzero $\l$. Moreover, $\K_{4}^{r,\non}=0$ as the fixed
torus is trivial.  We also expect that all $\K_{11}^r$ and $\K_{10}^r$
vanish from supersymmetry, as we will explicitly verify below.
Furthermore, one can see that all contributions from Eq.~(\ref{con0})
as well as from the curvature term in Eq.~(\ref{con2}) vanish: they
are proportional to $\str P_L$, which is just the sum over bosonic
minus fermionic degrees of freedom, weighed by their orbifold
phases. Since we have at least $\mathcal N=1$ supersymmetry
everywhere, this term vanishes for all sectors. The nonzero terms in
Eq.~(\ref{cont}) are thus $\K_{5}^{r,\non}$ and
$\K_{4,5}^{1,\loc}$. Writing $\l^{10}=2\pi n$, we have
\begin{align}
\K_{5}^{0,\non}&=\frac{\alpha_{5,0}}{6}(2\pi\s)
\sum_{n\neq 0} |2\pi n\s|^{-5}\str \left[\tfrac{1}{2}(2\pi n \,\omega_{10})^2(P^2_L+P_L^4)\right],\label{K31}\\
\K_{5}^{1,\non}&=\frac{\alpha_{5,1}}{6}(2\pi\s)
\sum_{n\neq 0} |2\pi n\s|^{-3}\str \left[(E+\omega^{\ell}\omega_{\ell})(P^2_L+P_L^4)\right],\label{K33f}\\
\K_{5}^{1,\loc}&=\frac{\alpha_{5,1}}{6}(2\pi\s)\Lambda_{UV}^3\str\left[(E+\omega^\ell\omega_\ell)(P^2_L+P_L^4)\right],\label{K33l}\\
\K_{4}^{1,\loc}&=\frac{\alpha_{4,1}}{6}\Lambda_{UV}^2\str\left[(E+\omega^\ell\omega_\ell+\omega^{10}\omega_{10})(P^1_L+P_L^5)\right],\label{K63l}
\end{align}
where $\ell=4\dots9$ and the constants $\alpha_{d,r}$ have been
defined in Eq.~(\ref{Stot}). The symbol $\str$ denotes the supertrace.
In the following we will calculate these terms and also verify the
cancellations for the sectors with $\mathcal N\geq 4$
supersymmetry. It is convenient to define the following combinations
of kinetic terms
\begin{align}
\O_1&=\textstyle\sum_{I<J=1}^3 \partial_\mu \log\rho_I\,\partial^\mu \log\rho_J\\
\O_2&=\textstyle\sum_{I=1}^3 (\partial_\mu \log\rho_I)^2\,,\\
\O_3&=\partial_\mu \log \sigma\textstyle\sum_{I=1}^3 \partial^\mu\log\rho_I\,,\\
\O_4&=(\partial_\mu\log\sigma)^2\,.
\end{align}

For the contribution $\K_{5}^{0,\non}$ we need to evaluate the supertrace
over the spin connection in Eq.~(\ref{K31}).  The spin connection along
the fixed torus $S^1$ is given by
\be
\omega_{10}=
-\Sigma^{10\,\beta}\partial_\beta \sigma\,
\label{spin}
\ee
where the $\Sigma^{AB}$ are the $SO(11)$ generators. Let us define the quantity
\be
C_k^{AB,CD}=\operatorname{str} \Sigma^{AB}\Sigma^{CD} P^k_L\,.
\ee
For the 5d sector we are interested in calculating $C_2$ and $C_4$.
Since the generators in Eq.~(\ref{spin}) are in fact generators of
$SO(5)\subset SO(11)$, $P^k_L$ commutes with the $\Sigma^{AB}$ and we
can symmetrize in the two generators. Each representation of $SO(5)$
has a definite phase $p$ under the orbifold action. The quantity
$C^{ab}$ above is then given by
\be
C_k^{AB,CD}=C \delta^{AB,CD}\,,\qquad C_k=\sum_p p \sum_{r_p} (-)^{F}   C_{r_p}
\label{Dynkin}
\ee
where $r_p$ label the different $SO(5)$-representations of a given
parity $p$ and $C_{r_p}$ is the corresponding Dynkin index.  The
easiest way to calculate the Dynkin indices is to consider the $SO(2)$
helicity group, which is a subgroup of $SO(5)$. This choice has the
advantage that one can restrict to physical states only and discard
any unphysical and ghost states that have to cancel each other. The
Dynkin index of an $SO(2)$ representation of helicity $h$ is
simply\footnote{The factor of 2 arises from the normalization: the
$SO(5)$ vector representation has $C_5=2$ in the standard convention
for the generators.}  $C_h=2h^2$ and all one needs to know to evaluate
$C$ is which 4d fields have a given parity. According to
Tabs.~\ref{parities} to \ref{paritiesf} we obtain
\begin{multline}
C_{2}=2\biggl[4\cdot1+ 1\cdot(10+9\theta+9\bar\theta)\biggr.\\
    \biggl.-\frac{9}{4}\cdot(2+3\theta+3\bar\theta)
    -\frac{1}{4}\cdot (20+18\theta+18\bar\theta)\biggr]=\frac{27}{2}
\end{multline}
The result for $C_{4}$ is the same.  
%
%
It remains to be shown that in the 11d and 10d sectors there occur
cancellations, as required by $\mathcal N\geq 4$ supersymmetry.  The
spin connection now transforms in $SO(11)$ ($SO(10)$) but we can again
apply our trick of calculating the Dynkin indices from the $SO(2)$
subgroup. From the tables one finds
\begin{align}
C_{k=0}&=2\left[4\cdot1 + 1\cdot28 -\frac{9}{4}\cdot 8 -\frac{1}{4}\cdot 56\right]=0\\
C_{k=3}&=2\left[4\cdot1+ 1\cdot(12-16)-\frac{9}{4}\cdot(4-4)-\frac{1}{4}\cdot (28-28)\right]=0
\end{align}

Let us then turn to the kinetic terms generated by the moduli
dependence of the mass matrices. 
According to our discussion in Sec.~\ref{toroidal}, we can use the
tree-level equations of motion in the one-loop correction to the
effective action, since -- up to higher order terms in the loop
expansion parameter -- this simply corresponds to a field
redefinition.  For the background we are considering here, the
equations of motion simply read $R_{MN}=0$.\footnote{Had we been
interested in the kinetic terms for the moduli originating from the
gauge sector or the $B$ field we would have to take into account terms
proportional to the energy momentum tensor when using the equations of
motion.}  Notice that this procedure also takes care of any additional
Weyl rescalings arising at one-loop order.  On-shell, the only nonzero
mass matrices are
\begin{align}
E^{\p{2t} MN}_{{2,t}\p{MN} PQ}&=-2 R^{(M\p{P} N)}_{\p{M}(P\p{N} Q)}\\
E^{\p{a3} M N L}_{a3\p{MNL} PQR}&=
- 6 R^{[M\p{P} N}_{\p{M}[P\p{M} Q}
\delta^{L]}_{\p {L}R]}\\
E^{\p{a2} M N }_{a2\p{MN} PQ}&=
- 2 R^{[M\p{P} N]}_{\p{M}[P\p{M} Q]}\\
E_{3/2\p A B}^{\p{3/2}A}&=-\frac{1}{2}R^A_{\p ABMN}\gamma^{MN}
\end{align}
(Recall that the ghosts for the gauge symmetries of the antisymmetric
three-form contain two real antisymmetric two-forms).  Clearly, for
the 11d sector, the trace over any of these matrices is proportional
to $R$ and hence again vanishes by the equations of motion.  For the
10d sector, notice that any trace $\tr P_L E$ can generally be
written in terms of $R_{(10)}$. However, the equations of motion also
imply $R_{(10)}=0$ and there are no K\"ahler corrections, as required
by $\mathcal N=4$ supersymmetry.  For the 5d and 4d sectors,
notice that $P_L$ always acts trivially on the 4d indices. The
equations of motion then allow one to make the replacements
\be
R^\mu_{\ i\mu j}=-R^k_{\ ikj}\,,\qquad R^{\mu\ \nu}_{\ \nu\ \mu}=R^{i\ j}_{\ j\ i}.
\ee
It should be clear at this point why the use of the equations of
motion can drastically simplify the analysis. In particular, there are
no one-loop terms proportional to the 4d curvature scalar and hence no
additional Weyl rescalings are necessary. The curvature tensor with
all compact indices can be expressed as
\be
R^{\ell r}_{\p{\ell r}sk}=(\partial_\mu \log\bar \rho_\ell)(\partial^\mu \log\bar \rho_r) 
(\delta^\ell_{\ k}\delta^r_{\p rs}-\delta^\ell_{\ s}\delta^r_{\p rk})
\label{allcomp}
\ee
where $\bar \rho_4=\bar \rho_5=\rho_1$, $\bar \rho_6=\bar
\rho_7=\rho_2$, $\bar \rho_8=\bar \rho_9=\rho_3$ and
$\bar\rho_{10}=\sigma$.
It is now straightforward to evaluate the traces. One finds for the
5d sector
\begin{align}
\tr E_t P_L^{\,2}&=-18 \O_1\\
\tr E_{a3} P_L^{\,2}&=6\O_2\\
-2\tr E_{a2}P_L^{\,2}&=-36\O_1-12\O_2\\
-1/2 \tr E_{3/2}P_L^{\,2}&=6\O_2
\end{align}
There is an identical contribution from the element $P_L^{\,4}$ in the sum
Eq.~(\ref{orbnodw}). Adding all contributions, one finds
\be
\str E (P^{\,2}_L+P_L^{\,4})= -108\mathcal O_1\,,
\ee
%
%
%
In a similar manner, for the 4d sector one finds
\be
\str E (P^1_L+P_L^{\,5}) =108 \O_1+36\O_2+48\O_3
\ee
%

To evaluate the traces over the square of the spin connection occurring
in Eq.~(\ref{K33f}) to (\ref{K63l}), notice that $\omega_\ell$ is
given by
\be
\omega_{\ell}=-\Sigma^{a\beta}\delta_{ai}\partial_\beta\bar\rho_i\,,
\label{spin2}
\ee
with the index $i,\ell=4\dots 9$. The $\Sigma^{AB}$ in
Eq.~(\ref{spin2}) are now generators that are broken by the $Z_N$
action, which changes the evaluation of $C^{AB,CD}$.  Using the
$SO(11)$ commutation relations as well as the orbifold transformations
of the generators one can write
\be
C_k^{a\alpha,b\beta}=\delta^{\alpha\beta} C_k^{ab}\,,\qquad 
C_k^{ab}=i(1-P^{-1})^{-1\,a}_{\p{-1 a} c}\str \Sigma^{cb}P^{\,k}_L\,,
\ee
Again, this vanishes for the 10d sector ($c=b=10$). For the 5d
and 4d sectors, $\Sigma^{cb}$ is a generator of $SO(6)$ or $SO(7)$
respectively, and the trace projects onto the $U(1)$ generator of the
surviving $U(3)$ subgroup, so we can write
\be
\str \Sigma^{cb}P_L=\frac{i}{3}Q^{cb}\str Q P_L=\frac{i}{3}Q^{cb}\sum_p p\sum_{q_p} (-)^Fq_p\,.
\ee
where $Q=\Sigma^{12}+\Sigma^{34}+\Sigma^{56}$. The charges can be read
off from Tabs~\ref{parities} to \ref{paritiesf}.  Without loss of
generality we can symmetrize $C^{ab}$ in the two indices, so we
finally obtain
\be
C_k^{(ab)}=\delta^{ab} C_k 
\ee
With 
\be
C_1=-C_5=\frac{9\,i}{2\sqrt 3}\,,\quad C_2=C_4=4\,,
\ee
%
This concludes the evaluation of the traces in Eq.~(\ref{K31}) to
Eq.~(\ref{K63l}). The result is thus
\be
\K_5^{0,\non}=\frac{27 \zeta(3)}{32\pi^4}\sigma^{-2}\,\O_4\,,\qquad
\K_5^{1,\non}=\frac{\zeta(3)}{4\pi^4}\sigma^{-2}\left[
-\frac{9}{4}\,\,\O_1+\frac{1}{3}\mathcal O_2\right]\,,
\ee
\be
\K^{1,\loc}_5=\frac{1}{\pi}\Lambda_{UV}^3\,\sigma\left[-\frac{9}{4}\O_1
+\frac{1}{3}\O_2\right]\,,
\ee
\be
\K^{1,\loc}_4=\frac{1}{4\pi^2}\Lambda_{UV}^2
\left[9\O_1+3\O_2+4\O_3\right]\,.
\ee
As in the previous subsection, one can recovariantize these terms in
order to make manifest the higher-dimensional invariances preserved by
the orbifolding. To this end, one should identify the $SO(4)\times
U(3)$ and $SO(5)\times U(3)$ singlets that one can form from the
curvature tensor and use them to replace the operators
$\O_i$.\footnote{ The $SO(5)\times U(3)$ singlets are, in the usual
complex basis, $\mathcal C_1=R^{IJ}_{\p{IJ}IJ}$ and $\mathcal
C_2=R^{I\bar J}_{\p{IJ}I\bar J}$. For $SO(4)\times U(3)$ one can, in
addition, form the invariant $\mathcal C_3
=R^{10\,I}_{\p{10I}10\,I}$. All other possible invariants are either
related to these by the equations of motion or by the symmetries of
the curvature tensor.  One can then immediately verify that $\mathcal
C_1\sim \mathcal O_1$, $\mathcal C_2\sim \O_1+\O_2$, and $\mathcal
C_3\sim O_3$. Note that the operator $\O_4$ originated from
the expansion of the Wilson line which, as a nonlocal object, does not
correspond to any local operator.} A direct evaluation of the
covariant result will be presented elsewhere \cite{future}.

\section{Conclusions}
\label{conclusions}

In this paper we have analyzed the one-loop effective action on
orbifolds. We have shown how the evaluation of the heat kernel in each
sector of the orbifold can be reduced to the one for the corresponding
fixed torus with a shifted mass matrix. We have proposed a further
expansion of the heat kernel coefficients in powers of the lattice
vectors defining the tori, and explicitly evaluated the expansion of
the coefficient $a_1$ to second order. Our formalism is carried out
entirely in position space, avoiding KK decomposition and displaying
very clearly the separation between local (UV sensitive)
renormalization and nonlocal (UV-finite) one. The main results of the
paper can be found in Eqns.~(\ref{Sloc}), (\ref{Sfin}),
(\ref{torus0}), (\ref{a1}), and (\ref{a2}) for the torus, and
Eqns.~(\ref{heat3}), (\ref{orbnodw}), (\ref{proj}), and (\ref{orbdw})
for the orbifold.

To exemplify our methods we have calculated the effective potential in
6d gauge theory on $T^2/Z_N$ and the corrections to moduli kinetic
terms in 11d supergravity on $T^6/Z_3\times S^1/Z_2$. In particular,
the latter example shows how K\"ahler corrections can be computed in
orbifold compactifications. This is extremely useful as it allows one
to analyze the moduli effective potential in a way that is independent
on the supersymmetry breaking mechanism.

Our results are restricted to operators that do not contain extra
dimensional derivatives.  For some applications (e.g.~effective
operators involving KK modes, warped backgrounds or otherwise
nontrivial profiles) one might wish to study backgrounds including
such normal derivatives. While the evaluation of the toroidal heat
kernel can be straightforwardly extended to this
case,\footnote{Gravitational backgrounds depending on the
extra-dimensional coordinate will require to replace the straight
lattice vectors $\l$ by the corresponding geodesics.} the orbifold
heat kernel receives further corrections.  These can be computed along the
following lines. Notice that, as a consequence of
Eq.~(\ref{spacegroupfields}), the heat kernel coefficients occurring
in the renormalization at, say, the fixed point $x_f=0$ satisfy the
following identity
\be
\tr\, a_r(x,Px) (P_L\otimes P_\G)=\tr\, \bigl[a_0(x_f,x)a_r(x,Px)a_0(Px,x_f)\bigr] (P_L\otimes P_\G)\,.
\ee
The quantity in square brackets is a covariant object at the fixed
point, i.e., it transforms at the fiber at $x_f$ from both
sides. There exists thus a covariant Taylor expansion in the geodesic
distance from the fixed point that has as coefficients gauge-covariant
operators at $x_f=0$. After integrating over $x$, the powers in this
expansion are replaced by powers of $T$. In this way one can obtain a
fully covariant fixed point action that takes into account the
complete invariance surviving the local projection. The explicit
calculation and evaluation of the expansion will be left to future
research \cite{future}.

\section*{Acknowledgments}
I would like to thank D.~Hoover for useful discussions.

\appendix

\section{Mass matrices}

\label{masses}

In this Appendix we would like to summarize the background dependence
of the inverse propagators, or fluctuation operators, for fields of
various spins, see for instance
Refs.~\cite{Vassilevich:2003xt,Hoover:2005uf}. We work in Euclidean
spacetime with the following conventions. The Christoffel connection
is given by $\Gamma^M_{SN}=-\frac{1}{2}\partial^M g_{SN}+\dots$ and
the curvature by $R^M_{\p MNRS}=\partial_R\Gamma_{SN}^M-\dots$ The
covariant derivative is $D_M=\nabla_M-iA_M-i\omega_M$ with hermitian
gauge and spin connections, the latter being related to the
Christoffel connection by
$\omega_M=-\frac{1}{2}\Sigma^{AB}e_A^N\nabla_Me_{NB}$. The conventions
for the vector generators of the Lorentz group is $(\Sigma^{AB})^C_{\
D}=-i(\delta^{AC}\delta^B_D-\delta^{BC}\delta^A_D)$.

For bosonic fields, the inverse propagator $\mathcal F$ is obtained by
linearizing the equations of motion in the fluctuations around a
generic background. For fermions, one takes the absolute square of
that operator. For a suitable choice of gauge, $\mathcal F$ can be
cast in the form
\be
\mathcal F=-D^2+2i\,B^ND_N+iD_NB^N+E\,,
\label{fluctgeneral}
\ee
where the covariant derivative $D$ contains all background gauge and
spin connections, and $E$ and $B_M$ are matrices depending on the
background fields.  The parametrization of Eq.~(\ref{fluctgeneral}) is
such that, with $B_M$ and $E$ hermitian, $\mathcal F$ is
hermitian.  The matrices $B_M$ and $E$ can mix different fields, in
particular particles of different spin. Note that we can
formally redefine the connection and the mass matrix to absorb the
terms linear in the derivative:
\be
\mathcal F=-(D-iB)^2+(E-B^2)\,.
\label{canonical}
\ee
Whereas off-diagonal elements in $E$ are relatively easy to deal
with,\footnote{Note that, in calculating $\tr E$, the off-diagonal
terms in $E$ do not contribute and only show up at $\O (E^2)$
\cite{Hoover:2005uf}.}  a nontrivial $B$ poses a bigger challenge from
a computational point of view. For the examples in this paper, we will
restrict to backgrounds that have $B=0$.  In the following, we give
the mass matrices for gauge theory and gravity.

\subsection{Gauge Theory}
  
We take as quantum fields all particles with spin $\leq1$, but will
include a general gravitational background.  We use the following gauge
fixing function in $R_{\xi}$ gauge with $\xi=1$.
\begin{align}
\mathcal G=D_M\mathcal A^M+iG^\dagger \phi-iG^T\phi^* \,.
\label{gauge}
\end{align}
where $G^{\p bA}_{b}=T^A_{bc}\phi_0^c$.  Here, $\mathcal\phi$ and
$\mathcal A$ are the dynamical fields and $G$ and $D$ only contain
backgrounds. All covariant derivatives as well as field strengths,
curvatures etc.~are to be evaluated at the background.

For complex scalar particles $\phi$ one finds
\be
E_0= \partial_{\phi^*}\partial_{\phi} V +GG^\dagger+\eta R\,,
\label{scalar}
\ee
where $V$ is the scalar potential and $\eta$ is an arbitrary
constant. For minimally coupled fields we have $\eta=0$ while for
conformally coupled ones we have $\eta=(d-2)/4(d-1)$.  The second term
in Eq.~(\ref{scalar}) comes from the gauge fixing.  For fermions one
finds
\be
E_{1/2}=-\Sigma^{AB} F_{AB}+\frac{1}{4}\,R
\ee
with $F$ denoting the field strength of the gauge connection. The
overall result has to be multiplied by $-1$, $-1/2$, or $-1/4$ for
Dirac, Majorana or Weyl, and Majorana-Weyl fermions respectively.
Note that $E_{1/2}$ is a $2^{[d/2]}$ dimensional matrix.
For the gauge fields themselves, one finds
\begin{align}
E_{1\, M N}&=2i\,F_{MN}+G^\dagger G\,g_{ MN}+R_{ MN}
\,,\\
E_{1, gh}&=G^\dagger G\,.
\end{align}
The second equation is the mass matrix for the ghost, whose
contribution to the effective action has to be multiplied by
$-2$. Notice that the matrix $F$ acts in the adjoint representation.
The matrix $E$ actually contains off-diagonal mixing terms as
discussed above. They are given by
\be
\Delta\mathcal L
=\frac{1}{2}\,\phi^TG^*G^\dagger\phi+i\,\phi^\dagger (D_M G)\mathcal A^M
+{\rm h.c.}
\ee
but do not contribute at $\mathcal O(E^1)$.

\subsection{Gravity}

Including dynamical gravitational fields is more involved, as now a
generic background generates mixing terms linear in derivatives as
discussed after Eq.~(\ref{fluctgeneral}). For instance, a nonzero
background for the gauge field induces terms such as
\be
(B_M)_{N,(PQ)}\sim F_{M(P}\,g_{Q)N}\,,
\ee
that mixes spin-one and spin-two fluctuations. For the sake of
simplicity, we shall consider purely gravitational backgrounds, in
which case one finds $B_M=0$.  Although slightly less general, this
background allows us, e.g., to calculate the effective action of the
gravitational moduli. The gauge fixings for the various gauge
symmetries are taken as in Ref.~\cite{Hoover:2005uf}. The mass
matrices for the fields with spin $\leq 1$ can be taken from the
previous subsection.
The mass matrix of a rank-p antisymmetric tensor field is given by
\cite{Fradkin:1982kf}
\begin{multline}
E^{\p{ap} M_1\dots M_p}_{ap\p{M_1} N_1\dots N_p}=
p\, R^{[M_1}_{\p {M_1}[N_1}
\delta^{M_2}_{\p {N_2}N_2}
\dots
\delta^{M_p]}_{\p {N_p}N_p]}\\
- p(p-1)\,R^{[M_1\p{N_1} M_2}_{\p{M_1}[N_1\p{M_2} N_2}
\delta^{M_3}_{\p {N_3}N_3}
\dots
\delta^{M_p]}_{\p {N_p}N_p]}
\end{multline}
where the square brackets on the indices denote antisymmetrization.
There are $p'-$form ghosts of any $0\leq p'<p$ that are fermions
(bosons) for $p-p'$ odd (even) and that occur in multiplicities of
$p-p'+1$. Their contribution to the effective action has thus to be
multiplied by $(p-p'+1)(-)^{p-p'}$.
For a Rarita-Schwinger field (gravitino) one has
\begin{align}
E_{3/2\, A B}&=
\frac{1}{4}\,R\, g_{AB}- i\, R_{ ABMN}\Sigma^{MN}
\\
E_{3/2, gh}&=\frac{1}{4}\,R 
\end{align}
The first line corresponds to the spin 3/2 field, its contribution to
the effective action has to be multiplied by $-1/2$ ($-1/4$) for
Majorana or Weyl (Majorana-Weyl) fermions.  There are three spinor
ghosts in total. Having bosonic statistics, the result has to be
multiplied by $+3/2$ $(3/4)$.  The dimension of the matrices $E_{3/2}$
and $E_{3/2, gh}$ are $d\cdot 2^{[d/2]}$ and $2^{[d/2]}$ respectively.
For the symmetric traceless part of the graviton one finds
\begin{multline}
E^{\p{2t} MN}_{{2,t}\p{MN} PQ}=R\left[ \delta^{(M}_{\p M(P}\delta^{N)}_{\p NQ)}
-\left(\frac{4}{d^2}+\frac{1}{d}\right)g^{MN}g_{PQ}\right]
-2\, R^{(M\p{P} N)}_{\p{M}(P\p{N} Q)}\\
+\frac{4}{d}\,(R^{MN}g_{PQ}+R_{PQ}g^{MN})-2 \,R^{(M}_{\p M(P}\delta^{N)}_{\p NQ)}
\end{multline}
where the parenthesis on the indices stand for their
symmetrization. Furthermore, the canonically normalized trace part
and the fermionic vector ghosts give a contribution
\begin{align}
E_{2,s}&=\frac{d-4}{d}\,R\,,\\
E_{2,gh\, MN}&=-  R_{ MN}\,.
\end{align}
The ghosts contribute with a factor $-2$ to the effective action.  Let us remark
that there are also mass mixings between the tensor and scalar modes
of the metric \cite{Hoover:2005uf}.

\section{Coincidence limits}

In this section we would like to review DeWitt's recursive procedure
to calculate the coincidence limits of heat kernel coefficients and
their covariant derivatives,
\be
[a_{r;\dots}]=\lim_{x'\to x} a_{r;\dots}(x,x')\,,
\ee
where the dots stand for any combination of primed and unprimed
indices and the semicolon denotes covariant differentiation.  The
ansatz Eq.~(\ref{ansatzcurved}) is inserted in the differential
equation Eq.~(\ref{heat}) to derive the recursion relations
\begin{align}
\s_;^{\ M} a_{0;M}&=0\,,\label{a0}\\
\s_;^{\ M} a_{r;M} +r a_r&=  \Delta^{-1}
(\Delta a_{r-1})_{;\ M}^M-E a_{r-1}\,.
\label{recursion}
\end{align}
where $\Delta=\bar\Delta^{1/2} (gg')^{-1/4}$ is a biscalar (as opposed
to $\bar\Delta$ which is a biscalar density).  It is now easy to
derive expressions for the coincidence limits needed in the evaluation
of the local part of Eq.~(\ref{Gamma}).  This is done by taking
repeated covariant derivatives of Eqns.~(\ref{biscalar}),
(\ref{vleck}) (\ref{a0}) and (\ref{recursion}), making use of the
commutation relations for covariant derivatives, and taking
coincidence limits \cite{DeWitt}. We will first calculate the
quantities with derivatives w.r.t.~$x$ only, the ones w.r.t.~$x'$ can
then easily be derived from Synge's rule 
\be
[X_{\dots}]_{;M}=[X_{\dots;M}]+[X_{\dots;M'}]\,.
\ee
In particular, one can show that\footnote{An extensive discussion of the quantities $\sigma$ and
$\Delta$ as well as their derivatives can be found in
Ref.~\cite{Poisson:2003nc}.} 
\be
[\s_{;M}]=[\s_{;MNR}]=[\Delta_{;M}]=[a_{0;M}]=0\,.
\ee
Recall that the boundary condition Eq.~(\ref{initial}) implies $[a_0]=\one$.
Hence,
\begin{align}
[a_1]&=[\Delta^M_{;\ M}+a^{\ M}_{0;\ M}-E]\,,\label{1}\\
2[a_{1;S}]&=[\Delta^M_{;\ MS}+a^{\ M}_{0;\ MS}-E_{;S}]\,,\\
3[a_{1;(ST)}]&=[\Delta^{-1}_{\ ; (ST)}\Delta^M_{;\ M}+\Delta^M_{;\ M(ST)}+\Delta_{;(ST)}a^{\ M}_{0;\ M}
    +\Delta^M_{;\ M}a_{0;(ST)}+\nn\\
    &\qquad +4\Delta^M_{;\ \l}a_{0;M\l}+a^{\ M}_{0;\ M(ST)}-E_{;(ST)} ]\label{3}\,,\\
2[a_2]&=[\Delta^M_{;\ M}a_1+a^{\ M}_{1;\ M}-E a_1]\,,\label{4}
\end{align}
and so on. Notice that $[a_{1;(ST)}]$ is needed both for the
evaluation of $[a_{2}]$ as well as $[a_{1;\l'\l'}]$ which enters in
Eq.~(\ref{a1}). The covariant expansion introduced in
Sec.~\ref{toroidal} is thus quite economic in that most of the algebra
for $[a_{1;\l'\l'}]$ is the same as for $[a_2]$.  To evaluate
Eq.~(\ref{1}) to (\ref{4}) one needs to know the coincidence limits of
$a_0$ and $\Delta$ with two, three and four derivatives which are
again obtained by differentiation of Eq.~(\ref{biscalar}),
Eq.~(\ref{vleck}) and Eq.~(\ref{a0}). The result is expressed in terms
of field strength and curvature tensors:
\begin{align}
[a_{0;MN}]&=-\tfrac{1}{2}\Omega_{MN}\,,\\
[a^{\ M}_{0;\ MN}]&=-\tfrac{1}{3}\Omega^M_{\p M N;M}\,,\\
[a^{\ M}_{0;\ M (ST)}]&=\tfrac{1}{2}\Omega_{ M(S}\Omega^M_{\p MT)}
    -\tfrac{1}{2}\Omega^{\p{M(S;}M}_{ M(S;\p M T)}+\tfrac{7}{12}R_{ M(S}\Omega^M_{\p M T)}\,,\\
[\Delta_{;(ST)}]&=\tfrac{1}{6}R_{ST}\,,\\
[\Delta^M_{;\ MS}]&=\tfrac{1}{6}R_{;S}\,,\\
[\Delta^M_{;\ M(ST)}]&=\tfrac{3}{20}R_{;ST}+\tfrac{1}{20}\Box
R_{ST}-\tfrac{1}{15}R_{MS} R^M_{\p M T}+\tfrac{1}{30}R^{MN}R_{MSNT}\nn\\
&\qquad +\tfrac{1}{36}RR_{ST}+\tfrac{1}{30}R^{MNL}_{\p{MNL}S}R_{MNLT}
\end{align}
where $\Omega$ is the field strength of the gauge and spin connections
\be
\Omega_{MN}=[D_M,D_N]=-iF_{MN}+\tfrac{i}{2}\Sigma^{AB}R_{ABMN}\,.
\ee
Inserting these expressions into Eqns.~(\ref{1}) to (\ref{4}) one gets
\begin{align}
[a_1]&=\tfrac{1}{6}R-E\,,\label{1'}\\
[a_{1;\l}]&=\tfrac{1}{12}R_{;\l}-\tfrac{1}{6}\Omega^M_{\p M \l;M}-\tfrac{1}{2}E_{;\l}\,,\\
[a_{1;\l\l}]&=\tfrac{1}{60}\Box R_{\l\l} +\tfrac{1}{20}R_{;\l\l}-\tfrac{1}{3}E_{;\l\l}
     - \tfrac{1}{45}R^M_{\p M\l} R_{M\l}+\tfrac{1}{90}R^{MN}R_{M\l N\l}\nn\\
&\qquad +\tfrac{1}{90}R^{MNL}_{\p{MNL} \l}R_{MNL\l}
     +\tfrac{1}{12}R^M_{\p M \l}\Omega_{M\l}+\tfrac{1}{6}\Omega_{ M\l}\Omega^M_{\p M\l}
    -\tfrac{1}{6}\Omega^{\p{M\l;}M}_{ M\l;\p M \l}\label{3'}\,,\\
[a_2]&=\tfrac{1}{2}(\tfrac{1}{6}R-E)^2+\tfrac{1}{6}\Box(\tfrac{1}{5} R-E)-\tfrac{1}{180}R_{MN}R^{MN}\nn\\
&\qquad +\tfrac{1}{180}R^{MNLS}R_{MNLS}+\tfrac{1}{12}\Omega_{MN}\Omega^{MN}\,,\label{4'}
\end{align}
Finally, using Synge's rule, one finds
\begin{align}
[a_{1;\l'}]&=-[a_{1;\l}]+[a_1]_{;\l}\\
[a_{1;\l'\l'}]&=[a_{1;\l\l}]-2[a_{1;\l}]_{;\l}+[a_{1}]_{;\l\l}
\end{align}
leading to Eq.~(\ref{a1}).

\label{colim}

\section{Zeta Regularization}

\label{zeta}

In performing the proper time integration of the heat kernel,
zeta-function regularization-techniques are often used
\cite{Dowker:1975tf} (see also
Refs.~\cite{Vassilevich:2003xt,Elizalde:1994gf}).  In this scheme, one
exploits the fact that the zeta function
\be
\zeta(s)=\Tr (-D^2+E)^{-s} \label{defzeta}\,,
\ee
is UV convergent for $s>\frac{d}{2}$ and has an analytic continuation that is
regular at $s=0$. One then writes formally
\be
S_{\eff}=(-)^F\,\frac{1}{2}\sum \Tr \log (-D^2+E)=-
(-)^F\,\frac{1}{2}\lim_{s\to 0}\zeta'(s)\,,
\ee
and uses the relation
\be
\zeta(s)=\Gamma(s)^{-1}\int dT\, T^{s-1}K(T)\,,
\ee
to write the renormalized effective action as
\be
S_{\eff}=-(-)^F\,\frac{1}{2}\,\lim_{s\to 0}\frac{d}{ds}\left(\Gamma(s)^{-1}\int dT \,T^{s-1}\,\Tr K(T)
\right)\,.
\label{zetaheat}
\ee
All relations in Eqns.~(\ref{defzeta}) to (\ref{zetaheat}) are well
defined at large $s$ and at $s=0$ after analytic continuation.  Of
course, if the integral in Eq.~(\ref{zetaheat}) is UV finite for $s=0$
one just recovers the old expression, Eq.~(\ref{Gamma}), by means of
the expansion $1/\Gamma(s)=s+\O(s^2)$. 
IR divergences have to be treated separately. We do
this by introducing an explicit mass $\mu$ to all fields such that a
suppression factor of $\exp(-\mu^2 T)$ is present in the integrals.
Let us define $\nu=r+s-\frac{d}{2}$, then we can write the integral
appearing in Eq.~(\ref{zetaheat}) as 
\be
\int \frac{dT}{T^{1-\nu}} \exp\left[ -T \mu^2-\frac{\l^2}{4T}\right]
=2(2\mu^2)^{-\nu}x^\nu K_\nu(x)\,,
\label{integral}
\ee
where $K_\nu(x)$ are the modified Bessel functions of the second kind
and we have defined $x=|\l| \mu $.  

For the nonlocal contributions, $\l\neq 0$, the integral is convergent
at $s=0$. Using that $1/\Gamma(0)=0$ and $[1/\Gamma(0)]'=1$ one finds
\be
S_{\eff,\,\non}=-(-)^F2^{-r-\frac{d}{2}}\pi^{-\frac{d}{2}}\sum_{r,\l\neq 0}
\left(\frac{|\l|}{\mu}\right)^{r-\frac{d}{2}} K_{r-\frac{d}{2}}(|\l|\mu)
\,[\bar\Delta]_\l\tr[a_r]_\l\,.
\ee
Being both IR and UV finite, this result is valid for all $r$ and $d$.
For $r<\frac{d}{2}$, the integration over $T$ is IR convergent and we
can take the limit $\mu\to0$. Using the small-x asymptotic expansion
\be
K_{r-\frac{d}{2}}(x)=K_{\frac{d}{2}-r}(x)\sim 2^{\frac{d}{2}-r-1}
\Gamma(\tfrac{d}{2}-r)x^{r-\frac{d}{2}}
\ee
we precisely recover Eqn.~(\ref{Sfin}).  However, the summation over
$\l$ is still IR sensitive as soon as the dimension of the operator
exceeds 4. It is reassuring that the presence of the IR cutoff also
takes care of the divergences for large $\l$ as the Bessel functions
are exponentially suppressed at large argument. Similarly, for
$r=\frac{d}{2}$ one uses
\be
K_0(|\l|\mu)\sim -\log(|\l|\mu')\,\qquad \mu'=\frac{e^{\gamma_E}}{2}\mu
  \sim 0.89\, \mu\,,
\ee
leading to
\be
S_{\eff,\,\non}^{r=d/2}=(-)^F(4\pi)^{-\frac{d}{2}}\sum_{\lambda\neq0} \log(|\l|\mu')
\int d^dx\,[\bar\Delta]_\l\tr[a_{\frac{d}{2}}]_\l\,.
\ee
%

For $|\l|=0$, the local contribution, the integral Eq.~(\ref{integral})
diverges at $s=0$ for $r<\frac{d}{2}$. Applying the prescription of analytic
continuation from the large $s$-region one finds
\be
\lim_{x\to 0}x^\nu K_\nu(x)=2^{\nu-1}\Gamma(\nu)\,.
\label{div}
\ee
Since the zeta function, Eq.~(\ref{defzeta}), and its derivative are
analytic at $s=0$ we expect that the poles of the Gamma function in
Eq.~(\ref{div}) cancel with the pole of the Gamma function in
Eq.~(\ref{zetaheat}). Let us define
\be
\beta_{d,r}(\mu/Q)\equiv  \frac{1}{2}(4\pi)^{-\frac{d}{2}}\,
\lim_{s\to 0}\frac{d}{ds}\left((\mu/Q)^{-2s}
\frac{\Gamma[r+s-\frac{d}{2}]}{\Gamma[s]}\right)\,,
\ee
where a renormalization scale $Q$ has been introduced to account for
the correct dimension. It can immediately be verified that this
quantity is finite for any $r$. Explicitely, one finds
\be
\beta_{d,r}(\mu/Q)=\frac{1}{2}(4\pi)^{-\frac{d}{2}}\left\{
\begin{array}{ll}
\frac{(-)^\nu}{(-\nu)!}\left[-\log\left(\frac{\mu^2}{Q^2}\right)+H_{-\nu}\right]&
\nu=r-\frac{d}{2}\leq 0,\ d\ {\rm even}\\
\Gamma(\nu)&{\rm else.}
\end{array}
\right.
\ee
Here, $H_n$ are the harmonic numbers defined as $H_n=\sum_1^n k^{-1}$
with the convention $H_0=0$.  The local part of the OLEA thus reads
\be
S^{\zeta \rm -reg}_{\eff,\loc}=-(-)^F\int d^dx \sqrt g\sum_r\beta_{d,r}(\mu/Q)\mu^{d-2r}\,\tr[\alpha_r]\,.
\ee
As with dimensional reguralization, zeta function reguralization does
not capture power-like divergences but only logarithmic ones (if
present).

\end{document}